\documentstyle[prl,aps,epsfig]{revtex}         % 2 col mode
\begin{document}
\draft               % preprint mode
% 2 col mode:
\twocolumn[\hsize\textwidth\columnwidth\hsize\csname @twocolumnfalse\endcsname

\title{Spatial Solitons in Resonators}
\author{C.O.Weiss, G.Slekys,$^*$ V.B.Taranenko, K.Staliunas}
\address{Physikalisch-Technische Bundesanstalt 38116 Braunschweig/Germany\\
$^*$present address: CNET, 92220 Bagneux/ France}
\author{R.Kuszelewicz}
\address{Centre National d'Etudes de Telecommunication,
Bagneux/France}
%\date{\today{}}
\maketitle
\begin{abstract}
We describe experiments testing the existence and investigating
the properties of spatial solitons in nonlinear resonators. We
investigate the properties of stationary and moving spatial
solitons in lasers with saturable absorber, with a subcritical
bifurcation, as well as their manipulation. As opposed, spatial
solitons relying on a supercritical bifurcation are shown to exist
in degenerate 4-wave mixing (DOPO). With a view to technical
applications in parallel information processing or communication,
experiments on spatial solitons in large area quantum well
semiconductor resonators are conducted.
\end{abstract}
\pacs{PACS 42.65.Sf; 42.65.Tg; 42.70.Nq} \vskip1pc ]
% BEGIN TEXT HERE

\section{Introduction\protect\\} \label{intro}

Solitary structures can form in optics when a balance occurs
between a linear and a nonlinear optical process. The best known
instance of such structures in optics are solitary pulses
propagating along optical fibers. Here a balance occurs between
the linear mechanism of dispersion, which leads to a broadening of
the pulse in the propagation direction, and the nonlinear
mechanism of "self-phase-modulation" or intensity-dependent
refractive index, (e.g. Kerr effect) which tends to shorten the
pulse. The result is a pulse traveling along the fiber without
changing its shape. Such pulses can well be described by the
soliton solutions of the 1 + 1D nonlinear Schroedinger equation
(NLSE) \cite{tag:1}. In this equation the time co-ordinate may as
well be a spatial co-ordinate. It follows that spatial solitons
should also exist. The linear broadening mechanism in this case is
diffraction. In one spatial dimension such spatial- or propagation
solitons do indeed exist \cite{tag:2}. The phenomenon manifests
itself in the contraction of a light beam with propagation, until
a filament of constant thickness is formed which then propagates
without further change ("Self-trapped beam").

Taking, however, a normal laser beam which diffracts or contracts
in 2D (the beam cross section), the NLSE has no stable solutions
of the form of a beam propagating with a constant diameter, at
least in the paraxial optics approximation. The Kerr-nonlinearity
is "stronger" than the diffraction, so that catastrophic collapse
of the beam cross section occurs \cite{tag:3}. It was pointed out
than such collapse can be avoided if the nonlinearity is
"saturable" i.e. reduces with increased light intensity
\cite{tag:4}. In this way a variety of experiments on propagation
solitons in 2D has been possible \cite{tag:5}. Technical
applications in information routing and field steering with such
propagation solitons are under consideration.

A particular situation occurs if such a "self trapped" beam
propagates inside an optical resonator. The finite mirror
reflectance acts in his case somewhat similarly to the saturation
of the nonlinearity because in each round-trip of the light, to
the already self-focused light unfocused light, which irradiates
the resonator, is added, thus continually weakening the
self-focusing. Consequently, in resonators of finite finesse
stable filamentation is possible \cite{tag:6}.

Evidently, the stability of such structures can be enhanced
further by a saturability of the nonlinearity of the material
filling the resonator. Thus, occurrence of spatial resonator
solitons has been predicted for a number of nonlinear materials
\cite{tag:6}. The first observations of such solitary structures
in optical resonators occurred before the bulk of theoretical work
on passive resonator solitons appeared. In \cite{tag:7} such
solitary structures were observed using a liquid crystal film
inside a resonator and in \cite{tag:8} such a spatial soliton was
observed in a resonator made up of two phase-conjugating mirrors
which contained a saturable absorber.

Spatial resonator solitons can exist if the characteristic of the
resonator shows two coexisting stable steady states (bistability).
In such a bistable resonator, if it is of large Fresnel number,
domains of the two states can exist, which are then connected by
"switching fronts" (or -waves). Such switching waves move into or
out of domains of one of the states depending on the difference of
the background field value and the field value corresponding to
the unstable steady state solution lying between the two stable
steady states. A switching front will move into the state of
higher intensity if the background field is larger than that of
the unstable steady state, and it will move into the domain of the
lower intensity state if the background field is smaller than that
of the unstable steady state. Thus in general one kind of domain
will shrink and the other expand.

In general the asymptotic state will be that the total resonator
cross section is entirely switched to one of the two states. If,
however, the system is not far from a modulational instability,
then the switching fronts do not aperiodically connect the two
states but can be accompanied by damped spatial oscillations on
either side of the fronts. Then as a domain contracts, finally the
switching front on one side of the domain will "feel" the spatial
oscillations of the field close to the front on the other side of
the domain. The spatial field minima can then "trap" the front of
the other side of the domain, in which case the (small) domain has
attained a stability (and is then called a solitary or localized
structure). It can be freely moved around the resonator cross
section. If the spatial field oscillations on the one side of the
domain trap not the front from the other side itself, but the
spatial oscillations accompanying it, then a higher order spatial
soliton is formed \cite{tag:9}.

We have found so far, that the solitons of low orders resemble
Gauss-Laguerre-Modes with ring nodes (not the flower-like
variety). There is no obvious reason why that should be so.
Optical resonator modes are the eigenfunctions of a boundary
problem with the boundaries given by the mirror surfaces.
Conversely, the solitons are the solutions of a self-consistence
problem where the "potential", constituted in case of an optical
resonator by the resonator mirrors, is created by the light field
itself. Thus it is not obvious why these two problems should have
similar solutions, and reasons for the similarity of the solutions
are an open question. Interestingly it has been found, that the
potential created by a fundamental soliton can actually
allow-besides the existence of the field of the fundamental
soliton -  the stable additional existence of a 1$^{st}$ order
soliton \cite{tag:10}.

One can, in all cases, picture a spatial soliton as a small domain
of one of two coexisting states surrounded by a stationary
switching front which has locked into a stable ring. Due to the
bistable character of the resonator such spatial solitons are
bistable. They can be switched on or off and are thus suitable for
carrying information.\\

\section{Solitons in laser with nonlinear absorber\protect\\} \label{sec:level1}

The theoretical work on solitons of active resonators dates back
to the 80s (see a summary in \cite{tag:11}).

It suggested experiments with a repetitively pulsed dye-laser with
an internal saturable absorber \cite{tag:12}. FIG. 1 shows the
output power of the dye laser with an internal Bacterio-Rhodopsine
(BR) absorption cell as a function of pump power. The pulse
repetition rate is 12 Hz, the acidity of the BR-absorber solution
is chosen for an absorber recovery time constant of ~300 ms so
that the system dynamics is slaved by the absorber and the system
can be treated like a continuously emitting system. The
bistability of the system is apparent. The upper part of FIG. 1
shows the output beam cross section. Apparently the narrowest beam
occurs within the bistable region. It represents a spatial
soliton.

\begin{figure}[htbf]
\epsfxsize=60mm \centerline{\epsfbox{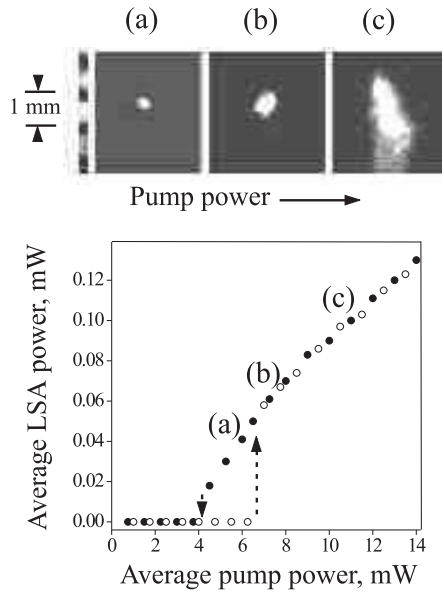}}
\vspace{0.5cm} \caption{Average output power of laser with
BR-absorber as a function of pump power  showing bistability.\\
Beam profiles at the positions  (a), (b), (c) indicated are given
above.}
\end{figure}

The resonator used is of "self-imaging" type. This kind of
resonator is in its transverse mode structure equivalent to a
plane resonator of zero length. For the precise self-imaging
length, its transverse modes are completely degenerate. The
diffraction losses, equally, correspond to a plane resonator of
zero length. Thus, this resonator, on the one hand, has the
complete transverse mode degeneracy of a plane resonator of zero
length as necessary for arbitrary images to resonate, and on the
other hand it has sufficient length to house various intracavity
elements without the detrimental diffraction losses of a plane
resonator of the same length. (It may be noted that such a
resonator permits to realize also a negative length as far as the
transverse mode structure concerns).

FIG. 2 shows the "writing" of a spatial soliton in this system in
various places of the resonator cross section. The absorber cell
was locally bleached for a short time (by a He-Ne laser beam). The
result is a stationary spatial soliton (which remains after the
external bleaching is stopped). FIG. 2 shows on the one hand that
the solitons are bistable (i.e. can be switched on and off) and on
the other hand that they can exist at any location in the cross
section.

The motion of solitons in field gradients was also tested. In a
fluid analogy of the laser \cite{tag:13} a phase gradient
corresponds to flow velocity and an intensity gradient to a
density- (or pressure-) gradient. Thus a soliton should move in
such gradients. FIG. 3 shows experiments with phase gradients. In
FIG. 3a a phase gradient across the laser cross section was
created by a small tilt of one laser resonator mirror. The
snapshots taken at equidistant times show the motion of the
soliton induced by the phase gradient. By changing the length of
the self-imaging laser resonator away from the precise-self
imaging length in FIG. 3b  a "phase trough" was created with its
minimum at the center of the resonator. As the snapshots show, the
soliton is drawn from all sides towards the center of the phase
trough, were it is then trapped. This movement and trapping of
solitons would likely be important for uses of resonator solitons
in optical information processing \cite{tag:14}.

\begin{figure}[htbf]
\epsfxsize=63mm \centerline{\epsfbox{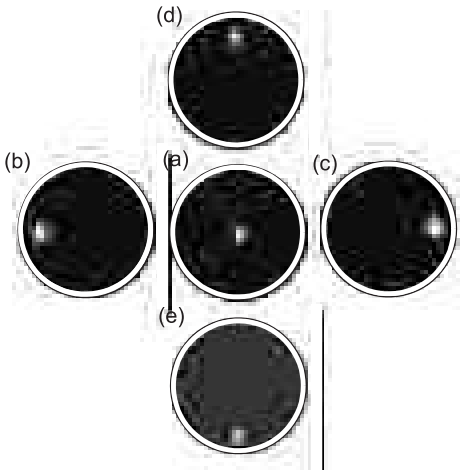}}
\vspace{0.5cm} \caption{A soliton can be "written" at any location
in the laser cross section.}
\end{figure}

In these initial experiments, the laser containing the nonlinear
absorber was emitting a larger number of longitudinal modes. Thus
the tuning of the resonator was of no importance, as the modes
emitted adjust to the resonator length. The resonator tuning does,
however, affect the solitons, their motion, and characteristics if
emission is restricted to a single longitudinal mode family. The
simplest way for such mode selection is an active medium with a
very narrow gain spectrum. By far the narrowest "gain" spectra (if
one interprets in a laser-physics concept) have photorefractive
gain media \cite{tag:15}. Therefore experiments were conducted
using resonators of self-imaging type \cite{tag:16} with
photorefractive gain. To picture the effects of a narrow gain line
one can think of the self-imaging resonator as a plane-plane
resonator. Such a resonator has a (longitudinal) resonance if the
length can accommodate an integer number $N$ of (half) wavelengths
of radiation generated - whose wavelength is fairly strictly given
by the wavelength of the center of the gain line.

Then, if the length of the resonator is between $N·\lambda/2$ and
$(N-1)·\lambda/2$ the emission can adjust to this resonator length
by tilting the propagation direction of the radiation with respect
to the resonator axis. In this way $N$ half wavelengths can be
accommodated in the resonator \cite{tag:17}. Such a "tilted wave"
has a propagation component along the resonator axis and a (small)
propagation component lying in the resonator mirror plane. Light
will therefore move across the resonator section in a detuned
resonator, corresponding to the motion in a phase gradient. We can
therefore expect stationary solitons in the case of precise
resonator tuning, and moving solitons for detuning.

\begin{figure}[htbf]
\epsfxsize=75mm \centerline{\epsfbox{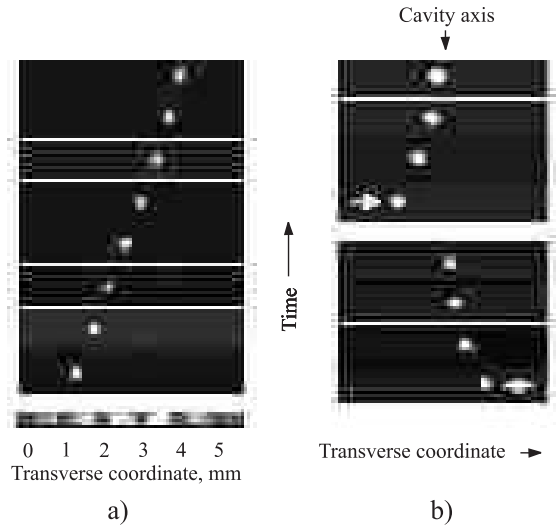}}
\vspace{0.5cm} \caption{A phase gradient (tilt of one resonator
mirror) leads to drift of the soliton (a)), a soliton drifts from
all directions into a phase trough (b)).}
\end{figure}

The stationary solitons, correspond to the stationary solitons
described above for the broadband (dye) multi mode laser which can
adjust its detuning to zero by choice of different longitudinal
modes. For the moving solitons corresponding to detuned (tilted
wave-)emission, one can from this picture directly deduce the
characteristics of the soliton motion:

1)  the direction of the tilt of the wave vector of the light
generated is free. Only the decomposition of the wave vector into
a longitudinal and a transverse component, in magnitude, is fixed.
Thus the direction of motion of the solitons is a priori
undetermined. The actual direction of motion is determined by
spontaneous symmetry breaking (in the same way as a single mode
laser choses the phase of its field at laser threshold). As for
the laser phase, there is here no restoring force for a particular
direction. The direction of motion of a soliton can therefore
change under external influence. In presence of noise it will
change diffusively.

2)  Whereas the direction of motion of a soliton, as well as its
position in the laser cross section are free, the magnitude of the
soliton velocity is fixed and given by the detuning of the
resonator (wave vector tilt is proportional to detuning).

3) Different longitudinal orders can be emitted simultaneously if
their wave vectors are tilted by different amounts. Thus in a
resonator of length N·$\lambda$/2, which emits a stationary
soliton, simultaneous emission of moving solitons is possible.
According to the different longitudinal orders, the velocities of
the moving solitons are quantized for a resonator of given length.

4) If we consider that in the experiment a self-imaging resonator
is used, with the nonlinear absorber in the near field (near a
plane mirror) and the (photorefractive) gain medium in the far
field, a strange type of competition among moving solitons
follows. Fields of two solitons overlapping inside the gain medium
compete. Only one soliton can then survive. The consequence is a
competition of solitons in $velocity$ $space$. If and only if two
solitons have the same vectorial velocity (i.e. velocity direction
and magnitude equal), they will compete. Notably, even when they
are far apart in the near field plane.

5) For stationary solitons the competition condition is trivially
fulfilled. Therefore only one stationary soliton can exist at a
time.

6) Moving solitons of different direction of motion and/or
different magnitude of velocity can coexist, thus in general a
large number of moving solitons can coexist with one stationary
soliton.

\begin{figure}[htbf]
\epsfxsize=65mm \centerline{\epsfbox{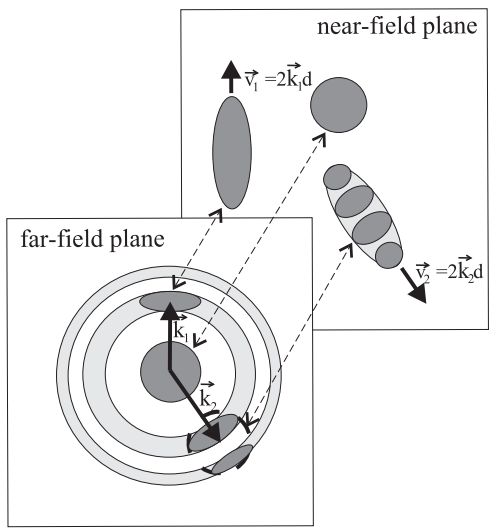}}
\vspace{0.5cm} \caption{Intensity distributions in the near and
far field for stationary and moving solitons. A central spot in
the far field corresponds to a stationary soliton with arbitrary
position in the near field. A spot (elliptical shape) on a
resonant ring corresponds to a moving soliton. Simultaneous
emission on two rings corresponds to a moving soliton  with
(stationary) modulation of the soliton intensity ("inch-worm").}
\end{figure}

FIG. 4 pictures the situation: a stationary soliton corresponds to
emission along the resonator axis. Moving solitons correspond to
emission at a fixed angle (given by detuning), to the resonator
axis, in the far field. Therefore the stationary soliton
corresponds to light in the central spot of the Airy rings of the
resonator, while the moving solitons correspond to emission on the
rings.

The restriction on the wavelength of the light generated is given
by the finite widths of the gain line of the active medium. The
finite width of this gain line corresponds to the allowed spread
of tilt angles of the emitted wave. Therefore the emission of the
stationary soliton corresponds to a central disk of finite
diameter and that of the moving solitons to finite area sections
of a ring. For moving solitons, the wavevector of emitted light
changes faster with radial angle than with azimuthal angle.
Therefore the light of a moving soliton in the far field occupies
an elliptical section of an Airy ring (see FIG. 4). Its Fourier
transform (moving soliton in the near field) is of elliptical
shape, with the long axis into the direction of motion (suggesting
again a fluid picture) FIG. 4.

FIG. 5 shows cases of all these soliton types as recorded on a
photorefractive BaTiO$_{3}$ oscillator with a BR-saturable
absorber which uses a self-imaging resonator \cite{tag:16}.
Excitation of emission segments on two rings into the same
azimuthal angle results in an "inch-worm"-soliton FIG.5.

\begin{figure}[htbf]
\epsfxsize=74mm \centerline{\epsfbox{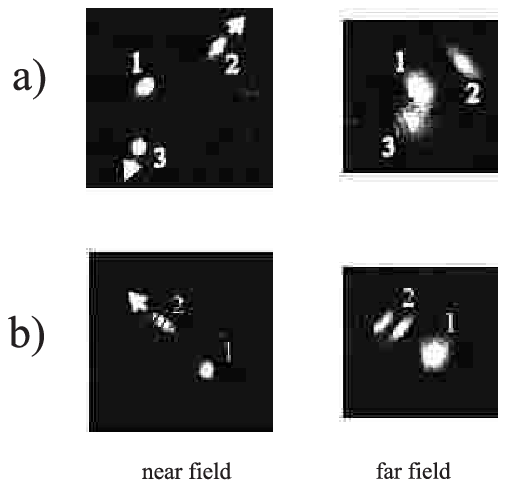}}
\vspace{0.5cm} \caption{Experimental realisation of the solitons
described in FIG. 4 (left: near field, right: far field):\\ a) one
stationary soliton 1 and two moving solitons 2, 3; \\ b) one
stationary soliton 1 and a moving "inchworm"-soliton 2.}
\end{figure}

In these experiments the gain elements of the lasers were placed
in the conjugate plane of the near field. This leads to
competition among certain solitons and, in particular, only one
stationary soliton can exist. For applications where such solitons
are to serve as binary elements for information storage, however,
large numbers of stationary solitons are desirable.

In order to test whether this is achievable, experiments were
conducted with the gain element and the nonlinear absorber both in
the near field plane. This case had been extensively treated in
\cite{tag:11} in the form of both elements being inside a plane
resonator. In the experiments the unsaturated absorption of the
nonlinear absorber was so high that, even at the highest pump
power available, the laser could not be brought to emit. External
bleaching of the absorber (by a green laser) was used for complete
saturation of the absorber so that large area laser emission
occurred. Reducing then the pump strength, the absorber gradually
unsaturates and becomes intensity dependent (nonlinear).

FIG. 6 shows the formation of the solitons in the experiment. To
illustrate the development of solitons out of the laser emission,
FIG.7 shows a numerical calculation of the process:

\begin{figure}[htbf]
\epsfxsize=26mm \centerline{\epsfbox{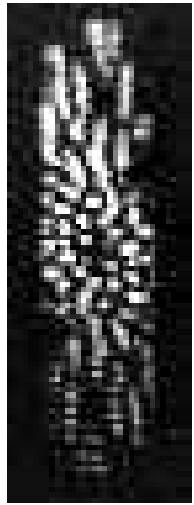}}
\vspace{0.5cm} \caption{Transition from 1D to 2D solitary
structures.}
\end{figure}

a) shows the emission typical for a tuned laser: A number of
optical vortices exist which are separated by "shocks"
\cite{tag:18} ("vortex glass" \cite{tag:19}).

b) as the pump is reduced, the vortices develop into dark areas
and the shocks convert to 1-dimensional soliton-structures (see
c), d)).

c)  Further reduction of the pump leads to shortening of the 1-D
solitary structures, which can then be converted to 2-D bright
solitons by increasing the pump slightly.

d)  This final increase in pump is necessary since the diffraction
losses for a 2-D soliton are larger than for a 1-D solitary line.
For details see \cite{tag:16}. This can be seen in FIG. 6: in the
outer regions where the pump is weak, stripes prevail, while at
the higher pump in the center spots dominate.

FIG. 8 finally shows ensembles of 2-D solitons in the final stage,
for different pump powers. It appeared that the number of solitons
existing in the final state is a monotonic function of the pump
power.

\begin{figure}[htbf]
\epsfxsize=52mm \centerline{\epsfbox{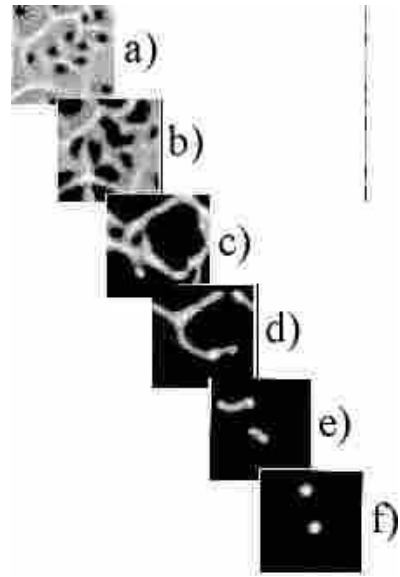}}
\vspace{0.5cm} \caption{Numerical calculation of the transition
from laser-to soliton emission.}
\end{figure}

\begin{figure}[htbf]
\epsfxsize=40mm \centerline{\epsfbox{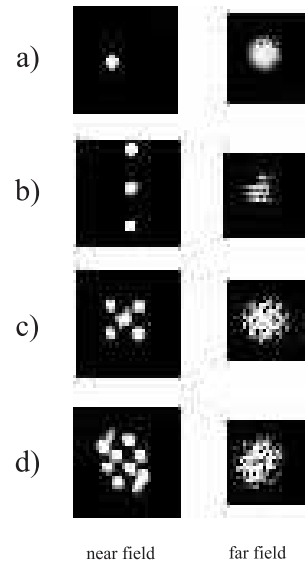}}
\vspace{0.5cm} \caption{Collections of solitons as experimentally
observed corresponding to different pump powers}
\end{figure}

As the pump beam has a Gaussian intensity profile, gradients
existed in the emission, which caused a slow outward motion of the
solitons. Thereby some solitons would reach the edge of the
emission field and extinguish there. This loss of solitons at
constant pump power was accompanied by continual splitting of
solitons in the central area of the near field. It appeared that
the splitting occurred to balance the loss of solitons at the
edges. FIG. 9 shows such soliton splitting.

\begin{figure}[htbf]
\epsfxsize=35mm \centerline{\epsfbox{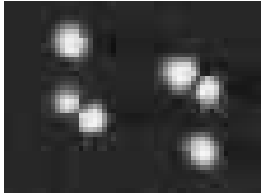}}
\vspace{0.5cm} \caption{Splitting of solitons in a laser with
saturable absorber (for details see text).}
\end{figure}

Time has not yet allowed us to study the interaction of solitons
in this system, which must exhibit interesting phenomena. Each of
the solitons here is an independent laser whose phase is
arbitrary. The interaction between solitons would on the one hand
depend on the relative phase, on the other hand two solitons whose
phase is free to change can be expected to synchronize their
phases. Whether this would be in or out of phase, combined with
the initial independence of the phases of the individual solitons
should produce a complicated interaction.

\section{Parametric mixing solitons\protect\\} \label{sec:level1}

Whereas the field of a laser, as used in the experiments described
above, can have any phase value, in wave-mixing with phase
matching the phase of the generated field is tied to the phase of
the pump field. The generated field can therefore be described as
a real-valued variable - as opposed to the complex-valued field of
a laser.

Spatial resonator solitons require, as has been described in Sec.
\ref{intro}, a bistable characteristic of the resonator.
Experiments described so far utilize a subcritical bistability
with a high- and a low-intensity branch. For degenerate wave
mixing such as 4-wave mixing (D4WM) or degenerate parametric
mixing (DOPO) a phase bistability of the (real-valued) field
occurs \cite{tag:20}. Although this is a symmetric and
supercritical bistability, one would expect that also this kind of
bistability would support spatial solitons, which we would call
$phase$-$solitons$. A calculation shows \cite{tag:21} that indeed
spatial solitons exist for a finite small detuning.

FIG. 10 gives shapes of such solitons in intensity and phase.
Inside the solitons the field phase is opposite to the surrounding
so that a dark circular interference fringe forms the switching
front connecting the two steady states.

The corresponding experiment was conducted using D4WM in
BaTiO$_{3}$ \cite{tag:22}. FIG. 11 shows the  resonator used. Two
pump beams together with the generated fields form an index
grating in the material which diffracts pump radiation into the
generated fields and adjusts self-consistently to the generated
field.

\begin{figure}[htbf]
\epsfxsize=60mm \centerline{\epsfbox{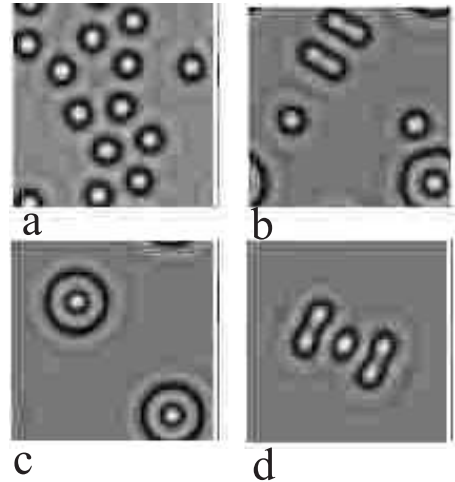}}
\vspace{0.5cm} \caption{Different stable localized structures
calculated for a DOPO or for D4WM. a) fundamental solitons, b)
bound states between fundamental solitons, and between fundamental
and higher order solitons c) higher order solitons, d) a
complicated bound state of solitons. }
\end{figure}

\begin{figure}[htbf]
\epsfxsize=85mm \centerline{\epsfbox{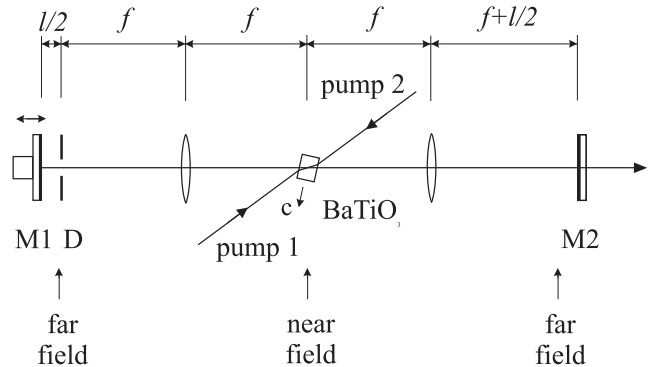}}
\vspace{0.5cm} \caption{Scheme of the resonator used for observing
phase-solitons in degenerate (photorefractive) 4-wave-mixing.\\ M:
mirrors, $\it{f}$: local length of lenses, $\it{l}$: deviation
from self-imaging length, D: iris for blocking high resonant
rings.}
\end{figure}

The two counter-propagating generated fields resonate in the same
(linear) resonator which forces their degeneracy and with that the
bistability and real value of the generated field. A typical
intensity distribution of the field generated experimentally is
shown in FIG. 12. Small circular domains coexist which larger
domains and black domain walls as was expected. FIG. 13 shows a
domain wall of complicated shape together with an interferogram
proving the opposite phase of the field on either side of the
domain wall. We note that the domain walls themselves are extended
1-D solitary structures \cite{tag:23}. They are the switching
waves connecting the two steady states of the resonator (field
with +$\pi$/2 and -$\pi$/2 phase). In general such switching waves
move and the domains they surround grow or shrink, the length of
the domain walls expending or contracting. The
expansion/contraction can be controlled by resonator detuning
\cite{tag:21}. FIG. 14 shows the contraction of a domain wall.

\begin{figure}[htbf]
\epsfxsize=40mm \centerline{\epsfbox{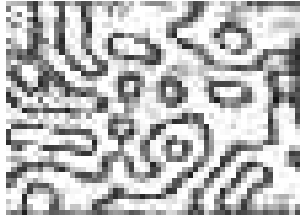}}
\vspace{0.5cm} \caption{A typical intensity distribution as
observed from a resonator as in FIG. 11. Small circular domains
coexist with large, irregularly shaped domains.}
\end{figure}

\begin{figure}[htbf]
\epsfxsize=75mm \centerline{\epsfbox{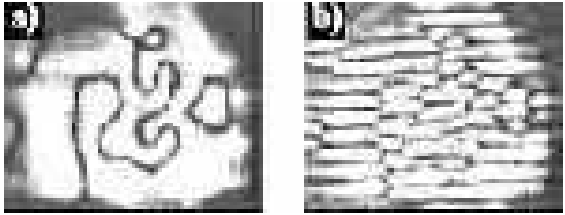}}
\vspace{0.5cm} \caption{A domain boundary of complex shape.
Interferogram shows that the fields separated by the boundary have
opposite phase (experimental).}
\end{figure}

Although in the experiment the small domains appeared to be
stable, it is necessary to prove their stability more explicitly
since stability is hard to distinguish from a slow transient
dynamics. Recordings under well-defined resonator tuning
conditions were therefore analysed. The resonator length was for
this purpose actively stabilized with respect to the pump light
frequency in a manner similar to that described in \cite{tag:24}.

\begin{figure}[htbf]
\epsfxsize=75mm \centerline{\epsfbox{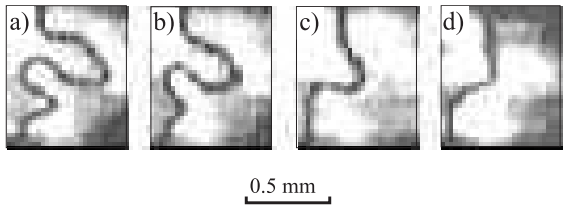}}
\vspace{0.5cm} \caption{Contraction of a domain boundary
(experimental).}
\end{figure}

FIG. 15 shows three snapshots out of an evolution captured in 20
frames. Fig.16 shows the change of the length of the boundaries of
the domains "1", "2", "3" as a function of time as measured on the
20 recorded frames. The largest domain-"1"-boundary shrinks
fastest. The medium sized domain-"2"-boundary shrinks at a slower
rate, while the domain-"3"-boundary does not change in time. This
is proof that the small circular domain "3" is stable and
represents a phase-soliton. The faster shrinking of the larger
domain is what is expected theoretically \cite{tag:21}.

\begin{figure}[htbf]
\epsfxsize=75mm \centerline{\epsfbox{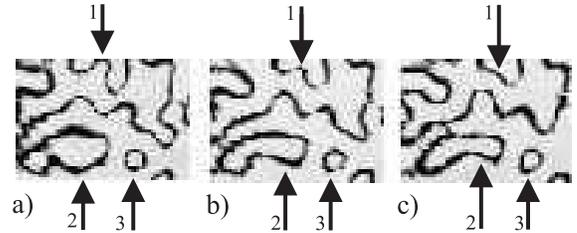}}
\vspace{0.5cm} \caption{Large domains 1, 2 shrink, while the small
domain 3 keeps constant size (experimental).}
\end{figure}

\begin{figure}[htbf]
\epsfxsize=55mm \centerline{\epsfbox{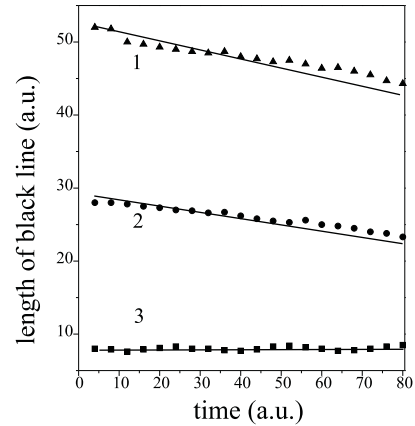}}
\vspace{0.5cm} \caption{Lengths of boundaries 1, 2, 3 from FIG. 15
as a function of time (experimental).}
\end{figure}

Stability of a soliton, under conditions where the shrinking rate
of a large domain is even larger, is shown in the four snapshots
FIG.17.

\begin{figure}[htbf]
\epsfxsize=75mm \centerline{\epsfbox{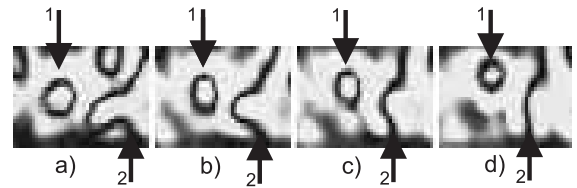}}
\vspace{0.5cm} \caption{Length of boundary 2 shrinks rapidly while
length of boundary 1 is constant in time (experimental).}
\end{figure}

Analysing the 20 frames out of which the FIG. 17 snapshots are
taken leads to FIG. 18 proving again than the small domain is a
phase-soliton.

\begin{figure}[htbf]
\epsfxsize=55mm \centerline{\epsfbox{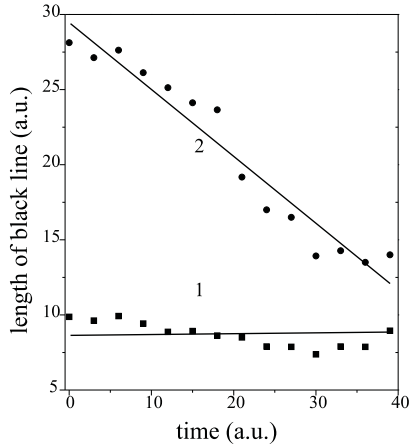}}
\vspace{0.5cm} \caption{Boundary lengths of FIG. 12 as a function
of time (experimental).}
\end{figure}

\section{Nonlinear semiconductor resonators\\} \label{sec:level1}

An interesting nonlinear material for technical applications is a
semiconductor. Solitons in semiconductor resonators were predicted
in \cite{tag:25}.

\begin{figure}[htbf]
\epsfxsize=70mm \centerline{\epsfbox{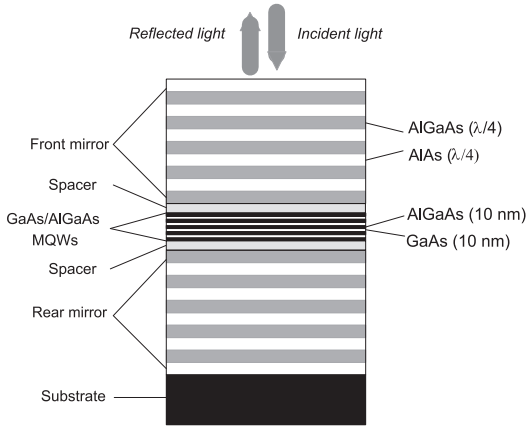}}
\vspace{0.5cm} \caption{Scheme of a semiconductor microresonator.}
\end{figure}

We have used a nonlinear semiconductor Fabry Perot for initial
experiments on spatial solitons. The nonlinear medium consists of
three quantum wells (FIG. 19). These three wells have between them
and 99,5 $\%$ Bragg mirrors, spacers to make the space between the
mirrors equal to a few $\lambda$/2. The thickness of this
structure is a few micron while the cross section of the resonator
is a few cm. Details about these nonlinear Fabry Perots are given
in \cite{tag:26}. The resonance of the Bragg resonator is slightly
dependent on the location on the sample, so that by choice of the
area to be irradiated the wavelength of excitation of the
semiconductor material can be chosen to lie either in the
interband transition, between interband transition and exciton
line, on the exciton line, or above the exciton line. Typically we
work a few 10 nm above the exciton wavelength so that the
nonlinearity is largely dispersive and defocusing. The empty
resonator finesse is around 500. With the residual absorption of
the semiconductor material the resonator finesse is  $\approx$
100. A cw Ti:Al$_{2}$0$_{3}$-laser is used for the excitation.

To avoid thermal effects the observations are done during
radiation pulses of a few microseconds length which are repeated
every millisecond. To create the pulses acousto-optic modulators
are used. The radiation is focused into a spot size of 50 - 100
$\mu$m on the semiconductor resonator surface, the light reflected
from the sample is observed by a CCD camera or by a fast (2ns)
photodiode.

As has been theoretically predicted for such dispersively
nonlinear resonators, under irradiation structure forms. FIG. 20
shows that the structure is a hexagonal lattice as expected
\cite{tag:27}.

\begin{figure}[htbf]
\epsfxsize=45mm \centerline{\epsfbox{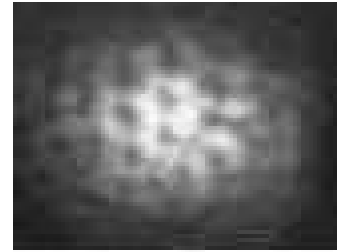}}
\vspace{0.5cm} \caption{Hexagonal patterns as observed on the
semiconductor nonlinear Fabry-Perot of FIG. 19.}
\end{figure}

Bistability of the resonator is easily reached (at intensities of
a few 100 W/cm$^{2}$). FIG. 21a shows the incident intensity
(dashed) and the reflected intensity (solid) as measured by the
fast photodiode.

The reduction of reflected light, as the sample is switched on, is
clearly seen, as is the increase of reflected light as the sample
is switched back off. From the intensities at which the switching
"on" and "off" occurs, the width of the bistability loop is
apparent. After the resonator was switched "on" at the point of
observation (image of detector) we varied the irradiating
intensity in order to observe the motion of the switching waves
connecting the on- and off-switched regions. By recording curves
as in FIG. 21a for different locations across a diameter of the
laser irradiation spot on the sample one is able to construct the
time history of the resonator dynamics on this diameter. FIG. 21d
shows a recording thus obtained.

Brightness in FIG.21d corresponds to reflectivity value. The
corresponding irradiation is given in FIG.21b in the form of
equi-intensity lines.

As the irradiation intensity is initially increased the
switching-on threshold is reached at a certain time in the center
of the laser field. A switching wave travels then outward until it
becomes stationary. We call it then "switching zone". As mentioned
in Sec. \ref{intro} a switching wave moves into the unswitched
region if the background intensity is larger than that
corresponding to the unstable steady state solution on the
unstable branch of the S-shaped resonator characteristic and vice
versa. Thus the switching wave becomes stationary at a particular
intensity corresponding to a certain distance from the maximum of
the Gaussian laser beam. This is what we observe.

\begin{figure}[htbf]
\epsfxsize=65mm \centerline{\epsfbox{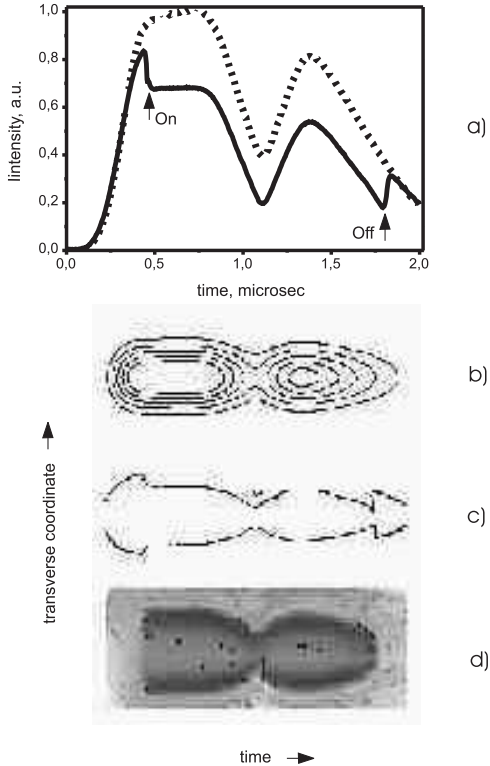}}
\vspace{0.5cm} \caption{Bistable switching of semiconductor
microresonator and spatial behaviour. a) Bistable switching: input
intensity: dashed line; reflected intensity: solid line. By
initial increase of irradiation the sample is switched "on" and by
decrease later "off".  In  between, intensity is varied to
demonstrate width of bistable region and to observe dynamics of
switching fronts. b) Equi-intensity lines (input intensity) for a
spatial co-ordinate on a diameter of the illuminated region, and
time. c) Motion of the switching front. d) Reflectivity of
sample.}
\end{figure}

When the power of the laser field is reduced, one would then
expect that the stationary switching wave (switching zone) would
move towards the center of the laser beam. Comparing  FIG. 21b and
FIG. 21d this is confirmed. The switching zone (boundary between
on- and off-switched areas) moves precisely on an equi-intensity
contour of the input light. FIG. 21c shows for clarity the
equi-reflectivity line corresponding to the switching zone, which
follows the second lowest intensity contour of the incident light.

If one chooses a location on the sample where the resonator
resonance is further from the exciton line, the switching zone
becomes accompanied on the lower branch side by fringes. This is
an indication that under these conditions the lower branch is
close to a modulational instability. This is a requirement for the
formation of spatial dark solitons as described in Sec. I.

\begin{figure}[htbf]
\epsfxsize=45mm \centerline{\epsfbox{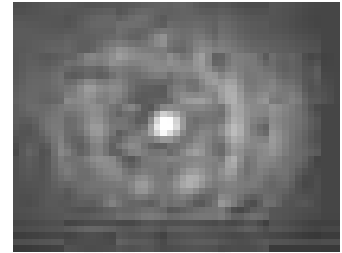}}
\vspace{0.5cm} \caption{A small (10 $\mu$m) bright structure
giving first indication of a soliton.}
\end{figure}

Suitable choice of parameters appears indeed to lead to a solitary
structure. FIG. 22 shows a bright narrow spot of the size the
order of the elements of the hexagonal pattern FIG. 20. (FIG. 22
is an average over 20 laser pulses with rectangular
intensity-vs-time-form). We can clearly show that this small
structure is bistable as predicted for a soliton: FIG. 23 shows
the proof. We use a rectangular laser pulse with a constant
intensity in the middle of the bistability region. A short
increase in intensity beyond the upper intensity of the
bistability switches the small localized structure "on". It
remains "on" until the intensity is for a short time reduced to
below the lower intensity of the bistability.

\begin{figure}[htbf]
\epsfxsize=65mm \centerline{\epsfbox{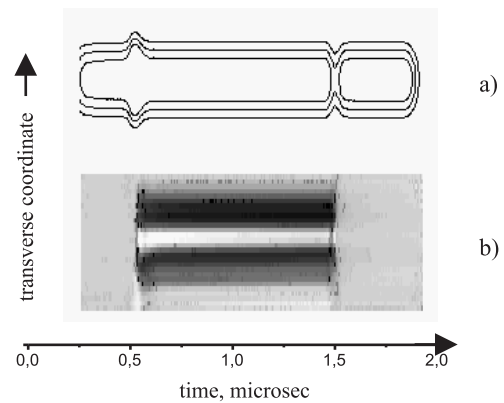}}
\vspace{0.5cm} \caption{Test of bistability of the small structure
of FIG. 22. a)  Equi-intensity lines of input light. b) Sample
reflectivity.}
\end{figure}

This clearly demonstrates that the small structure is bistable as
expected for a spatial soliton. We have tried to record that this
structure has a stability of its shape as one would expect for a
soliton (or a circularly locked switching zone). This is shown in
FIG.24.

\begin{figure}[htbf]
\epsfxsize=65mm \centerline{\epsfbox{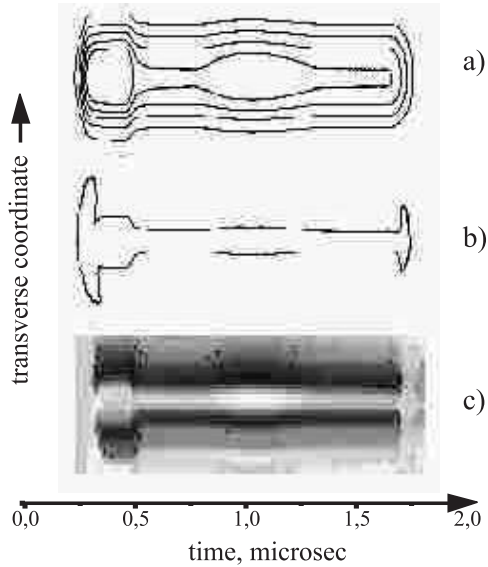}}
\vspace{0.5cm} \caption{Test of stability of the small bright
structure. a) Equi-intensity lines of input light. b)  Motion of
the switching front. c) Light intensity reflected from the
sample.}
\end{figure}

FIG. 24a gives the equi-intensity lines of the input field. During
a short high intensity period at the beginning, the central part
of the beam is switched up(see reflected intensity FIG. 24c).
Reducing the intensity lets the switched-up region then contract
to the small diameter of FIG. 22.

We test the stability of this structure now by a variation of the
light intensity: if the small bright structure is just a circular
switching zone (which is not locked and thus is not a soliton)
then its diameter should follow a contour of the incident light.
FIG. 24b shows the contour corresponding to the switching zone.
Evidently it does not follow any of the contours of the incident
light (FIG. 24a). This indicates a certain robustness of the
narrow structure against changes of system parameters, for which
reason FIG. 22 can be taken as the first indication of the
existence of localized structures in semiconductor resonators.

A more explicit test on the existence of such independent
localized structures was possible by injection of spatially
narrow, temporarily short light pulses into the illuminated area
\cite{tag:28}. FIG. 25a shows a collection of bright spots
resulting from illumination of the area shown. The narrow pulse is
first directed at the spot marked "a". As can be seen in FIG. 25b,
this switches the bright spot "a" to dark. All other spots
remaining unchanged.

Correspondingly, the second pulse was directed at bright spot "b"
which is equally switched to dark, all other spots remaining
unchanged, as seen in FIG. 25c. As the FIGs 25 a to c are
time-average-pictures not indicating directly the switching, FIG.
25d gives the intensity at the center of a switched spot as a
function of time. The upper trace corresponds to a pulse energy
not sufficient for switching and no permanent switch of the bright
spot results. With sufficient energy of the pulse, however,
permanent switching occurs, i.e. the intensity remains small
throughout the illumination (until the reduction of the background
intensity near the end of the illumination returns the resonator
to monostable.) Thus, in this case, individual bright spots are
found which can be independently of the rest of the system,
switched. This makes the bright spots observed very "soliton-like"
\cite{tag:28}.

\begin{figure}[htbf]
\epsfxsize=75mm \centerline{\epsfbox{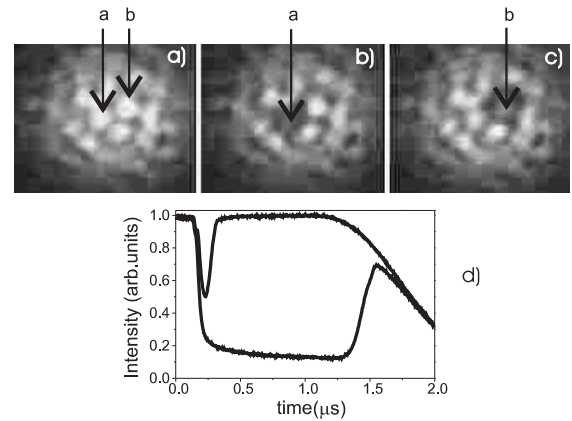}}
\vspace{0.5cm} \caption{Switching of individual bright spots of a
bright spot cluster (see text). Nonlinearity is defocusing
dispersive. }
\end{figure}

To experiment with single bright spots, in order to show their
soliton nature, was not possible under the largely dispersive
conditions used, because the collection of bright spots appears
largely as a consequence of linear filtering of the high finesse,
high Fresnel number resonator. For details see \cite{tag:28}.

In order to suppress this linear ("noise induced"-) structure, we
chose to work at lower resonator finesse i.e. closer to the band
edge or the exciton line, where, moreover, the dissipative
solitons predicted in \cite{tag:14} could be more likely expected.
FIG. 26 shows observations (pictures were taken as snapshots of 50
ns duration) under these conditions.

FIG. 26a shows a switched area (resonator field is high in the
dark area because observation is in reflection) surrounded by a
switching front. For small intensities such switched area
collapses into the structure shown in FIG. 26b, which shows all
the characteristic features of a bright soliton (dark due to
observation in reflection), particularly, the spatial oscillations
around it \cite{tag:29}.

We have recently been able to switch this structure on and off by
a narrow pulse similar to what was done to observe Fig. 25d.

Interestingly, at higher illumination intensity "dark" solitons
(bright in reflection) appear. FIG. 26c shows such a soliton;
embedded in the upswitched area as to be expected \cite{tag:29}
Such dark solitons were predicted in \cite{tag:30}. As predicted
there, we have found that these "dark" solitons are less stable
than the bright ones. They appear to move and we find that for
smaller intensities they tend to pulse in a regular fashion,
somewhat similarly to what was predicted in \cite{tag:30}.

\begin{figure}[htbf]
\epsfxsize=65mm \centerline{\epsfbox{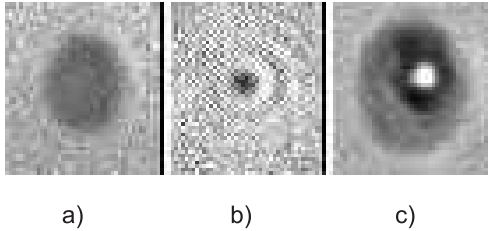}}
\vspace{0.5cm} \caption{Switched structures: reflectivity
(reflected light/incident light) of the sample: a) switched domain
(limited by a contour of Maxwellian intensity), b) bright soliton
(dark spot in reflection), c) dark soliton (bright spot in
reflection).}
\end{figure}

A hint towards the nonlinear nature of these structures comes from
the brightness of the light reflected on the structure.
Quantitative intensity measurement \cite{tag:29} shows that the
light reflected at the center of the structure is almost twice as
high as the illumination intensity. This means a reflectivity
higher than one. This has to be interpreted such that the
structure collects light from its surrounding and emits it at its
center.

\begin{figure}[htbf]
\epsfxsize=65mm \centerline{\epsfbox{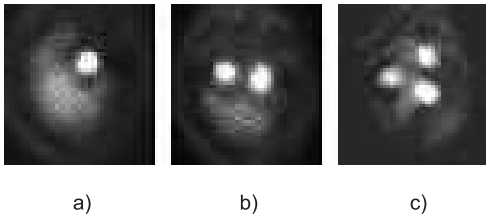}}
\vspace{0.5cm} \caption{Dark solitons observed in the switched
area (intensity).}
\end{figure}

FIG. 27 finally shows that more than one soliton can exist, even
at our conditions which are limited by finite laser power and
spatial nonuniformity of the illuminating field \cite{tag:29}.
With these solitary structures in semiconductor microresonators,
information processing and storage should be possible.\\

This work was supported by ESPRIT projects PASS and PIANOS. Growth
of semiconductor Fabry-Perots by I.Sagnes is gratefully
acknowledged.

\end{document}